\documentclass[aps,prl,twocolumn,groupedaddress,showpacs]{revtex4}
\usepackage{keyval}%
\usepackage{graphicx}
\usepackage{dcolumn}
\usepackage{bm}
\setkeys{Gin}{width=0.7\columnwidth,clip=true}

\newcommand{\ignore}[1]{}

\begin{document}

\title{Photoemission spectra of a two-dimensional S=1/2 quantum antiferromagnet in magnetic
fields: a theoretical study}

\author{Wei-Guo Yin}
\email{wgyin@yahoo.com}
\author{W. N. Mei}%
\affiliation{%
Department of Physics, University of Nebraska, Omaha, NE 68182}


\date{\today}

\begin{abstract}
We calculate the angular resolved photoemission spectra (ARPES) of
a spin-1/2 Heisenberg antiferromagnet as a function of magnetic
fields using both the exact diagonalization method and
self-consistent Born approximation. Below the saturation field
$B_C$, strong scattering between spin waves and a hole, created by
photoemission of an electron, significantly narrows the
quasiparticle band that is characterized by the lowering of the
quasiparticle energy at $(\pi,\pi)$ with increasing field.
Accordingly, in ARPES the quasiparticle peak gets shaper near 
$(\pi,\pi)$ and broader elsewhere. Furthermore, we observe that 
an anomalous extended van Hove region (EVHR) around $(\pi,\pi)$ 
appears in a half saturation field, while EVHRs around $(\pi,0)$ 
and $(0,\pi)$ in zero field gradually disappear with increasing field.

\end{abstract}


\maketitle



The effect of a strong magnetic field on quantum spin systems has
played an integral role in the understanding of magnetism and
quantum phase transitions. There is renewed interest in this topic
due to the synthesis of a new family of low-dimensional quantum
antiferromagnets with small exchange constants \cite
{AFM:dender,AFM:zhou,AFM:hammar,AFM:woodward}. For example, the
theoretical prediction of field-induced incommensurate soft modes
in one-dimensional (1D) spin-1/2 quantum antiferromagnet (AFM) was
confirmed by neutron scattering experiments on copper benzoate
\cite{AFM:dender}, while anomalous spin excitation spectrum of a
2D AFM in strong fields has been reported
\cite{AFM:lee,AFM:zhitomirsky-PRL}. Other important observations
include superconductivity in a layered organic AFM at very high
magnetic fields \cite{OSC:uji,OSC:balicas}. In this paper, we
present a theoretical study of the magnetic field dependence of
the angular resolved photoemission spectra (ARPES) in a 2D AFM.

Zero-field ARPES experiments in a 2D AFM are important to the
research of many cuprate superconductors whose undoped parent
compounds are nearly square lattice spin-1/2 antiferromagnetic
insulators \cite {ARPES:shen}. It was revealed that the presence
of strong scattering processes between spin waves and a hole,
created by emission of an electron, gives rise to an extended van
Hove singularity near the Fermi surface. This novel feature is
crucial in understanding many other anomalous physical properties
of the cuprate superconductors and could be explained in the
spin-polaron picture \cite {ARPES:shen98,yin:prl98}. On the other
hand, above the critical field $B_C$, the system becomes a
saturated ferromagnet and the ARPES are known to be free-particle
like. Therefore, it would be interesting to study the evolution of
the ARPES of a 2D AFM from one limit to the other, in particular,
the change of the extended van Hove regions. To our knowledge,
this problem remains unanswered, primarily because the exchange
constants in typical planar cuprates are $J\sim 1500$ K, hence,
only near zero-field studies were possible. Nevertheless, a newly
fabricated family of spin-1/2 square lattice antiferromagnets,
(5CAP)$_2$Cu$X_4$ and (5MAP)$_2$Cu$X_4$ with $X$=Cl or Br
\cite{AFM:zhou,AFM:hammar,AFM:woodward}, was found to have small
$J\simeq 0.57-8.5$ K with $B_C \sim 2-24$ T, thus provides a good
testing ground for our theoretical analysis. Alternatively, the
ARPES studies could be carried out in certain pseudospin systems
where the effective magnetic field $B_\mathrm{eff}$ is comparable
with the effective exchange constant $J_\mathrm{eff}$; for
instance, in a mixed valence system with electronic
ferroelectricity \cite{efe:batista,yin:xxx03}, $B_\mathrm{eff}$ is
the $d$- and $f$-level energy difference that could be as large as
$J_\mathrm{eff}$ and adjustable by alloying or applying pressure.
In the present work, we observe a magnetic-field-induced anomalous
flat region around $(\pi,\pi)$, where the bottom of the
quasiparticle band locates, in a magnetic field of
$\frac{1}{2}B_C$, signifying an extended van Hove singularity near
the Fermi surface from zero field up to $\frac{1}{2}B_C$.

Our starting point is a 2D $t$-$J$ model, a simple yet effective
Hamiltonian to model layered cuprate superconductors, in a
magnetic field along $z$ direction,
\begin{equation}
H=-t\sum_{\langle \mathbf{i},\mathbf{j}\rangle ,\sigma }\,(\widetilde{c}_{%
\mathbf{i}\sigma }^{\dagger }\widetilde{c}_{\mathbf{j}\sigma }^{}\,+%
\widetilde{c}_{\mathbf{j}\sigma }^{\dagger }\widetilde{c}_{\mathbf{i}\sigma
}^{})+J\sum_{\langle \mathbf{i},\mathbf{j}\rangle }\mathbf{\vec{\tau} }_{\mathbf{i}%
}\cdot \mathbf{\vec{\tau} }_{\mathbf{j}} - 
B\sum_{\mathbf{i} }\tau _{\mathbf{i}}^z,  \label{1}
\end{equation}
where $\widetilde{c}_{\mathbf{i}\sigma }^{}=c_{\mathbf{i}\sigma }^{}(1-n_{%
\mathbf{i}\overline{\sigma }})$ is the constrained fermion operator, and
$\mathbf{\vec{\tau}}_{\mathbf{i}}=\sum_{\mu \nu }\widetilde{c}%
_{\mathbf{i}\mu }^{\dagger }\mathbf{\vec{\sigma}}_{\mu \nu }^{}\widetilde{c}%
_{\mathbf{i}\nu }^{}$ with $\{\mathbf{\vec{\sigma}}_{\mu \nu
}^{}\}$ being the Pauli matrices is the spin operator. The
\textrm{SU(2)} symmetry is broken in the presence of an applied
magnetic field, for the AFM will orient itself in such a way that
the staggered direction, which is chosen to be the $x$ direction,
is perpendicular to the applied field. The resulting magnetic
phase is a canted state of two sublattices in which the spins tilt
towards the $z$ axis by the angle $\theta =\arccos (B/B_C)$ with
the saturation field $B_C=4J$. 
Note that 
$\theta =0$ for $ B \ge B_C$. The
Hamiltonian is invariant under rotation of spins around the $z$
axis, hence the spin excitation spectrum should be a gapless
Goldstone mode.

The ARPES are defined as
\begin{equation}
A(\mathbf{k},\omega )=\sum_\nu \,\left| \left\langle 1,v\left| {\tilde{c}_{%
\mathbf{k,}\sigma }}\right| 0\right\rangle \right| ^2\,\delta (\omega
-E_0+E_{1,v}),  \label{2}
\end{equation}
where $E_0$ and $|0\rangle $ are the ground-state energy and
eigenvector without any hole, and $E_{1,\nu }$ and $|1,\nu \rangle
$ are the energy and the wave vector of the $\nu$-th eigenstate
with one hole created by photoemission of one electron. To
calculate $A(\mathbf{k},\omega )$, we first employ the Lanczos
exact diagonalization (ED) algorithm \cite{HTC:dagotto} with $100$
iterations and an artificial broadening factor $\eta =0.05t$ on a
$4\times 4$ square cluster. Furthermore, in order to overcome the
finite size effect in the ED calculations and gain more insights,
we perform in the following analytical calculations of
$A(\mathbf{k},\omega )$ in the spin-polaron picture in the context
of the self-consistent Born approximation.

To simply the notation, it is convenient to perform a rotation of
the spins in the $A$ and $B$ sublattices by $\theta $ and $-\theta
$ about the $y$ axis, respectively:
\begin{eqnarray*}
\widetilde{c}_{\mathbf{i}\uparrow } &=&\cos \frac \theta 2d_{\mathbf{i}%
\uparrow }-e^{i\mathbf{Q\cdot r}_i}\sin \frac \theta 2d_{\mathbf{i}%
\downarrow },\; \nonumber \\
\widetilde{c}_{\mathbf{i}\downarrow } &=&e^{i\mathbf{Q\cdot r}%
_i}\sin \frac \theta 2d_{\mathbf{i}\uparrow }+\cos \frac \theta 2d_{\mathbf{i%
}\downarrow }, \nonumber \\
\tau _{\mathbf{i}}^x &=&S_{\mathbf{i}}^x\cos \theta +S_{\mathbf{i}}^ze^{i%
\mathbf{Q\cdot r}_i}\sin \theta,\; \nonumber \\
\tau _{\mathbf{i}}^z &=&S_{\mathbf{i}}^z\cos
\theta -S_{\mathbf{i}}^xe^{i\mathbf{Q\cdot r}_i}\sin \theta,
\end{eqnarray*}
where $\tau _{\mathbf{i}}^y=S_{\mathbf{i}}^y$ and $\mathbf{Q}=(\pi ,\pi )$. 
$ d_{\mathbf{i}\sigma }$ are constrained fermion operators and $%
\overrightarrow{\mathbf{S}}_{\mathbf{i}}=\sum_{\mu \nu
}d_{\mathbf{i}\mu }^{\dagger }\mathbf{\vec{\sigma}}_{\mu \nu
}^{}d_{\mathbf{i}\nu }^{}$ are spin operators in the new local
coordinate system. This canonical transformation maps the canted
spin configuration to a ferromagnetic configuration with all spins
up and removes further necessity to distinguish between
sublattices. The Hamiltonian thus has the form $H=H_t+H_J$, where
\begin{eqnarray}
H_t &=&-t\cos \theta \sum_{\langle \mathbf{i},\mathbf{j}\rangle }
(d_{\mathbf{ i}\uparrow }^{\dagger }d_{\mathbf{j}\uparrow }^{}+d_{\mathbf{i}\downarrow
}^{\dagger }d_{\mathbf{j}\downarrow }^{}) \nonumber \\
&\; &  -t\sin \theta \sum_{\langle \mathbf{i},\mathbf{j}\rangle } e^{i\mathbf{Q\cdot r}_i}
(d_{\mathbf{i}\uparrow }^{\dagger }d_{\mathbf{j}\downarrow }^{}-d_{%
\mathbf{i}\downarrow }^{\dagger }d_{\mathbf{j}\uparrow }^{})+H.c., \\
H_J &=&J\sum_{\langle \mathbf{i},\mathbf{j}\rangle }(S_{\mathbf{i}}^zS_{%
\mathbf{j}}^z\cos 2\theta +S_{\mathbf{i}}^xS_{\mathbf{j}}^x\cos 2\theta +S_{%
\mathbf{i}}^yS_{\mathbf{j}}^y) \nonumber \\
&\,&+J\sin 2\theta \sum_{\langle \mathbf{i},\mathbf{j}\rangle }(S_{\mathbf{i}%
}^zS_{\mathbf{j}}^x-S_{\mathbf{i}}^xS_{\mathbf{j}}^z) \nonumber \\
&\,&+B\sin \theta \sum_{\mathbf{i}} S_{\mathbf{i}}^xe^{i\mathbf{Q\cdot r}_i}
-B\cos \theta \sum_{\mathbf{i}}S_{\mathbf{i} }^z.
\end{eqnarray}
Then, we treat quantum spin fluctuations within linear spin wave
theory \cite{yin:xxx03,AFM:zhitomirsky}, namely, we regard the
deviations of the spins measured from their equilibrium directions
are small, hence only up to the quadratic terms of the deviation
operators are retained in $H_J$. Since the equilibrium directions
are determined from vanishing of the linear terms, the second and third terms
in $H_J$ that are linear in spin deviations could be neglected.
Note that there are spin-flipping hopping terms $\propto e^{i\mathbf{Q\cdot r%
}_i}(d_{\mathbf{i}\uparrow }^{\dagger }d_{\mathbf{j}\downarrow }^{}-d_{%
\mathbf{i}\downarrow }^{\dagger }d_{\mathbf{j}\uparrow }^{})$ in
$H_t$, which represent emitting and absorbing spin excitations,
thus contribute to strong scattering processes between the hole
and spin waves. Taking the ferromagnetic configuration as the
vacuum state, we employ the slave-fermion formalism to cope with
the constraint of no doubly occupancy
in $H_t$ \cite{HTC:SVR}. Defining holon (spinless fermion) operators $h_{%
\mathbf{i}}$ so that $d_{\mathbf{i}\uparrow }=h\,_{\mathbf{i}}^{\dagger }$,$%
\,\,d_{\mathbf{i}\downarrow }=h_{\mathbf{i}}^{\dagger }a_{\mathbf{i}}^{}$
where $a_{\mathbf{i}}=S_{\mathbf{i}}^{+}$ is the hard-core boson operator,
we arrive at an effective spin-polaron Hamiltonian in the momentum space
\begin{eqnarray}
H &\simeq & \sum_{\mathbf{k}}\varepsilon _{\mathbf{k}}^{}h_{\mathbf{k}}^{\dagger
}h_{\mathbf{k}}^{}
+\sum_{\mathbf{q}}\omega _{\mathbf{q}}^{}\alpha _{\mathbf{q}%
}^{\dagger }\alpha _{\mathbf{q}}^{} \nonumber \\
&\; & +\sum_{\mathbf{k},\mathbf{q}}(M_{\mathbf{kq}}h_{\mathbf{k}%
}^{\dagger }h_{\mathbf{k}-\mathbf{q}}^{}\alpha _{\mathbf{q+Q}%
}^{}+H.c.), 
\end{eqnarray}
where $\alpha _{\mathbf{q}}$'s are spin wave operators, $a_{\mathbf{q}}=u_{%
\mathbf{q}}\alpha _{\mathbf{q}}+v_{\mathbf{q}}\alpha _{-\mathbf{q}}^{\dagger
}$, $\,$with dispersion $\omega _{\mathbf{q}}^{}=(A_{\mathbf{q}}^2-B_{%
\mathbf{q}}^2)^{1/2}$. The transformation coefficients are $u_{\mathbf{q}%
}=[(A_{\mathbf{q}}/\omega _{\mathbf{q}}^{}+1)/2]^{1/2}$ and $v_{\mathbf{q}}=-%
\mathrm{sgn}(B_{\mathbf{q}})[(A_{\mathbf{q}}/\omega _{\mathbf{q}%
}^{}-1)/2]^{1/2}$. 
Here the shorthand notations are
$A_{\mathbf{q}}=2J(1+\cos ^2\theta \gamma _{\mathbf{q}})$, $B_{\mathbf{%
q}}=2J\sin ^2\theta \gamma _{\mathbf{q}}$, and $\gamma
_{\mathbf{q}}=(\cos q_x a+\cos q_y a)/2$ with the lattice constant
$a$ being the length unit in the following. Note that the
spin-wave spectrum is indeed gapless ($\omega _{\mathbf{q=Q}}=0$).
The bare hole dispersion is $\varepsilon _{\mathbf{k}}^{}=4t\cos
\theta \gamma _{\mathbf{k}}$, and the hole-spin-wave coupling is
\begin{equation}
M_{\mathbf{kq}}=\sin \theta \frac{4t}{\sqrt{N}}(\gamma _{\mathbf{k}-\mathbf{q%
}}u_{\mathbf{q+Q}}+\gamma _{\mathbf{k}}v_{\mathbf{q+Q}}).
\end{equation}
Using the self-consistent Born approximation (SCBA) in which the
spectral functions of one hole in a $t$-$J$-$like$ model can be
accurately calculated
\cite{yin:prl98,HTC:SVR,HTC:leung,HTC:bala,yin:prl01}, that is, we
first compute the hole Green's function $G(\mathbf{k},\omega
)=[\omega -\varepsilon _{\mathbf{k}}^{}-\Sigma (\mathbf{k},\omega
)+i0^{+}]^{-1}$ self-consistently with the self-energy,
\begin{equation}
\Sigma (\mathbf{k},\omega )=\sum_{\mathbf{q}}M_{\mathbf{kq}}^2G(\mathbf{k}-%
\mathbf{q},\omega -\omega _{\mathbf{q+Q}}).  \label{se}
\end{equation}
Thus, the spectral functions of the hole quasiparticle (QP) are
given by $A(\mathbf{k},\omega )=-\mathrm{Im}
G(\mathbf{k},\omega )/\pi $, the spectral weights are $Z(\mathbf{k}%
)=[1-\partial \Sigma (\mathbf{k},\omega )/\partial \omega ]_{\omega=E_%
\mathbf{k}}^{-1}$, and the QP dispersion is $E_{\mathbf{%
k}}\equiv \varepsilon _{\mathbf{k}}+\mathrm{Re}\Sigma (\mathbf{k},E_{\mathbf{%
k}})$.

\begin{figure}[tbp]
\includegraphics*{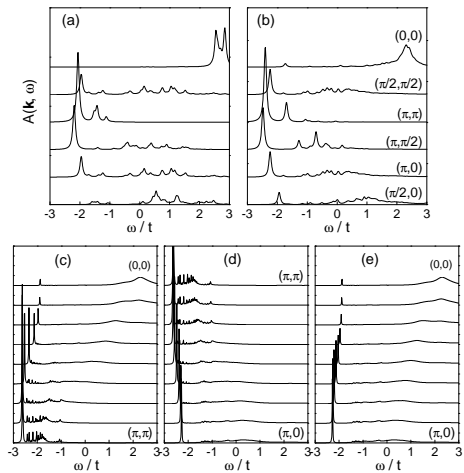}
\caption{\label{fig:arpes}%
The ARPES $A(\mathbf{k},\omega)$ of the $4 \times 4$
antiferromagnet for $J = 0.4t$ and $B=\frac{1}{2}B_C$ using (a) ED
and (b) SCBA. (c)-(e) present the ARPES of the $16 \times 16$
antiferromagnet along (c) $ (0,0)-(\pi,\pi)$, (d)
$(\pi,\pi)-(\pi,0)$ and (e) $(\pi,0)-(0,0)$ directions using SCBA.
}
\end{figure}

\begin{figure}[tbp]
\includegraphics*{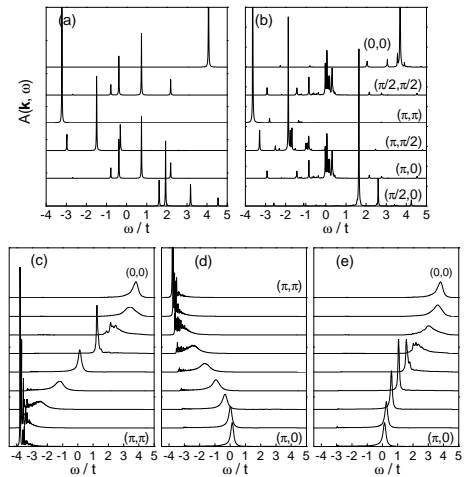}
\caption{ \label{fig:B14}%
Same as in Fig.~\ref{fig:arpes}, except for $B=\frac{7}{8}B_C$.}
\end{figure}

\begin{figure}[tbp]
\includegraphics*{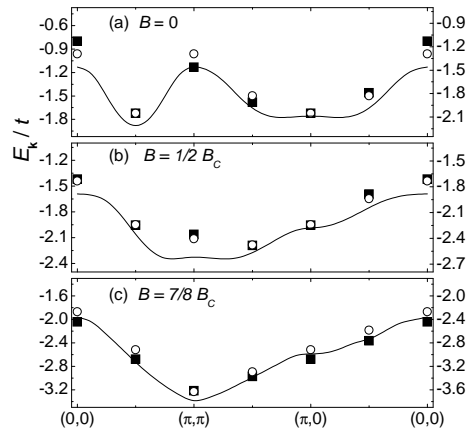}
\caption{\label{fig:ED}%
Quasiparticle dispersion for $J = 0.4t$ and $B=0$ (a),
$\frac{1}{2}B_C$ (b), and $\frac{7}{8}B_C$ (c). Left label: ED
results (solid squares) on the 4 x 4 square lattice. Right label:
SCBA on the 4 x 4 (open circles) and 16 x 16 (solid lines) lattices.%
}
\end{figure}

To examine the finite size effect, we start the SCBA calculations
with a $4\times 4$ square lattice and compare the results with
those obtained from using ED, then gradually increase the lattice
size. Finally, we find that the SCBA results obtained from a
$32\times 32$ lattice are rather close (within 1\%) to those from
a $16\times 16$ lattice. This implies that the SCBA results on the
$16\times 16$ lattice are not seriously affected by the boundary
effect. Since in typical cuprates, $J \sim 0.2-0.5 t$, we will
adopt $J=0.4t$ in our calculations. The energy mesh is chosen from
$-6t$ to $6t$ with interval $0.01t$.

Then we proceed to compare the ED and SCBA results. From the ARPES
$A(\mathbf{k},\omega)$ of the $4 \times 4$ AFM in
Figs.~\ref{fig:arpes}(a)-(b) for $B=\frac{1}{2}B_C$ and in
Figs.~\ref{fig:B14}(a)-(b) for $B=\frac{7}{8} B_C$, we notice that
the line shapes obtained from using ED [Figs.~\ref{fig:arpes}(a)
and \ref{fig:B14}(a)] agree well with those obtained from SCBA
[Figs.~\ref{fig:arpes}(b) and \ref{fig:B14}(b)]. As for the
results for $B=0$, we refer readers to
Ref.~\cite{HTC:dagotto,HTC:leung}. These numerics unambiguously
demonstrate that the spin-polaron picture provides indeed a
natural description of the QP behavior. In addition, the ARPES of
the $16 \times 16$ AFM obtained from using SCBA shown in
Figs.~\ref{fig:arpes}(c)-(e) and Fig.~\ref{fig:B14}(c)-(e) are in
consistence with those obtained from the $4 \times 4$ AFM. The
main feature of these ARPES is that as $B$ increases, the spectrum
at $(\pi,\pi)$, the QP band bottom, gets more and more coherent,
while away from the band bottom the spectra becomes incoherent
quickly, indicating that the QP states away from the minimum decay
by emission of spin waves.

Furthermore, we present in Fig.~\ref{fig:ED} the calculated
electronic structure of the 2D Heisenberg AFM with $B=0$,
$\frac{1}{2}B_C$ and $\frac{7}{8}B_C$, respectively. Again, the
SCBA results agree well with those obtained from ED. We notice
three main features in these quasiparticle bands: First, as $B$
increases, the band bottom evolves from $(\pi/2,\pi/2)$ to
$(\pi,\pi)$. Second, the low-field band [Fig.~\ref{fig:ED}(a)] has
flat regions around $(\pi, 0)$ and $(0,\pi)$, while the high-field
band [Fig.~\ref{fig:ED}(c)] resembles the free particle dispersion
but its width is severely narrowed. In between, for
$B=\frac{1}{2}B_C$ [Fig.~\ref{fig:ED}(b)], a new anomalous
feature, a flat region around $(\pi,\pi)$, appears in the
electronic structure, which leads to a strongly distorted density
of states with a massive peak near the bottom of the QP band.
Therefore, an extended van Hove singularity near the bottom of the
QP band survives up to $B=\frac{1}{2}B_C$. Third, the QP bandwidth
is strikingly narrowed, which can be seen more clearly in
Table~\ref{table.dat}.

In Table~\ref{table.dat}, we present several interesting physical
quantities, including the QP bandwidth $W$, the spin-wave
bandwidth $W_{\mathrm{sw}}$, the minimal QP energy
$E_\mathrm{min}$, and the QP spectral weights $Z(\mathbf{k})$ at
high symmetric points in the first Brillouin zone, as a function
of the magnetic field. These data are obtained from using SCBA on
the $16 \times 16$ lattice. We find that $W$ and $W_{\mathrm{sw}}$
have similar values. Such band narrowing can be understood as
follows: since gapless spin excitations are easily stimulated by
the incoherent motion of a hole, the combination of the hole and
polarized spin-wave cloud constitutes the quasiparticle,
spin-polaron. Hence, $W$ scales with $W_{\mathrm{sw}}$. Another
special feature revealed from Table~\ref{table.dat} is that in low
fields the largest spectral weights locate at $(\pi/2,\pi/2)$ and
$(\pi,0)$; increasing $B$ would raise $Z(\pi,\pi)$, while it
decreases the QP spectral weights at other wave vectors. This
confirms the spectral features shown in Figs.~\ref{fig:arpes} and
\ref{fig:B14}. We explain this feature as follows: in low fields
the spins are almost antiferromagnetically aligned, therefore
effective hole hopping in the same sublattice would be favored,
leading to $E_\mathrm{min}$ at $(\pi/2,\pi/2)$ and considerably
large values of $Z(\pi/2,\pi/2)$ and $Z(\pi,0)$. On the contrary,
increasing $B$ will raise the bare hole dispersion $\varepsilon
_{\mathbf{k}}^{} \propto \cos \theta = B/B_C $ and decrease the
hole-spin-wave coupling $M_\mathbf{kq} \propto \sin\theta$, thus
facilitates the hole propagation to nearest neighboring sites and
lowers the QP energy at $\mathbf{k}=(\pi,\pi)$, whereas hole
motion with other wave vectors is impeded due to scattering off
spin waves. We notice from Table~\ref{table.dat} that $W$ and
$Z(\mathbf{k})$ experience abrupt change as $B \rightarrow B_C$.
In this limit, Eq.~(\ref{se}) can be solved analytically in
perturbation theory, since $M_{\mathbf{kq}} \rightarrow 0 $ as $B
\rightarrow B_C$. Hence, the QP dispersion is given by
\begin{equation}
E_\mathbf{k} \simeq \varepsilon_\mathbf{k}
+\sum_{\mathbf{q}}\frac{M_{\mathbf{kq}}^2}{\varepsilon_\mathbf{k}-\varepsilon_\mathbf{k-q}-\omega
_\mathbf{q+Q}}=\varepsilon_\mathbf{k}+O(1-\frac{B}{B_C}).
\label{perturb}
\end{equation}
Above $B_C$, $M_{\mathbf{kq}}=0$, thus $E_\mathbf{k} = \varepsilon_\mathbf{k} =
4t\gamma _\mathbf{k}$, that is, the hole moves freely in a
saturated ferromagnet.

\begin{table}
\caption{\label{table.dat}%
$E_\mathrm{min}$, the minimal QP energy, $W$, the QP bandwidth,
$W_\mathrm{sw}$, the spin-wave bandwidth, and $Z(\mathbf{k})$,
the QP spectral weights at several high symmetric points in the first
Brillouin zone as a function of $B$, the magnetic field, obtained
on the $16\times 16$ lattice using SCBA for $J=0.4t$.}
\begin{ruledtabular}
\begin{tabular}{cccccccc}
$B/B_C$ &  $E_\mathrm{min}/t$  &   $W/t$   &   $W_\mathrm{sw}/t$  &   %
$Z(0,0)$  & $Z(\frac{\pi}{2},\frac{\pi}{2})$  &  $Z(\pi,\pi)$  &  $Z(\pi,0)$  \\
\hline
  0 &   -2.209  &   0.748   &   0.800   &   0.1112  &   0.3466  &   0.1112  &   0.3821  \\
1/8 &   -2.213  &   0.731   &   0.800   &   0.0580  &   0.3426  &   0.2047  &   0.3730  \\
2/8 &   -2.269  &   0.685   &   0.801   &   0.0406  &   0.3147  &   0.3068  &   0.3431  \\
3/8 &   -2.399  &   0.674   &   0.811   &   0.0304  &   0.2587  &   0.3946  &   0.2907  \\
4/8 &   -2.639  &   0.757   &   0.849   &   0.0214  &   0.1809  &   0.4698  &   0.2178  \\
5/8 &   -3.021  &   0.972   &   1.000   &   0.0121  &   0.0857  &   0.5300  &   0.1298  \\
6/8 &   -3.418  &   1.199   &   1.200   &   0.0076  &   0.0186  &   0.5891  &   0.0447  \\
7/8 &   -3.786  &   1.427   &   1.400   &   0.0013  &   0.0013  &   0.6550  &   0.0043  \\
1 &   -4      &  8    &   1.600   &   1   &   1   &   1   &   1    \\
\end{tabular}
\end{ruledtabular}
\end{table}

It is worth mentioning that linear spin-wave theory was pointed
out to be inaccurate in strong fields -- spin waves in a quantum
AFM in a strong field ($>0.76 B_C$) were unstable with respect to
the spontaneous two-magnon decays \cite{AFM:zhitomirsky-PRL}.
Nevertheless, we have demonstrated that the strong-field ARPES
obtained from using the SCBA, where the spin excitations are
treated in linear spin-wave theory, agree well with the ED
results. Hence, it appears that linear spin-wave theory is
appropriate to calculate the spectral functions of a hole in a
quantum AFM. While it is desirable to follow the present study
with high order spin-wave theory or ED for larger lattices, 
we anticipate that our basic conclusions remain valid.

\ignore{It is expect that the ARPES above a threshold field ($\sim
0.76 B_C$ \cite{AFM:zhitomirsky-PRL}) would be more coherent than
those presented here. We anticipate that high order spin-wave
theory would be able to address this problem and the calculations
are in progress.}

To summarize, we have studied the magnetic field dependence of the
ARPES of a square lattice spin-1/2 Heisenberg antiferromagnet.
Using the self-consistent Born approximation in the spin-polaron
picture, we observe an anomalous electronic structure in which an
extended van Hove region around $(\pi,0)$ in zero field moves
toward $(\pi,\pi)$ with increasing field strength until reaching
a half saturation field. We also present the exact diagonalization
data that confirm our prediction, which could be tested by ARPES
experiments on weakly interacting spin-1/2 square lattice
antiferromagnets \cite{AFM:zhou,AFM:hammar,AFM:woodward}.

We are grateful to C. D. Batista and H.-Q. Lin for useful discussions.
This work was supported by the Nebraska EPSCoR-NSF Grant
EPS-9720643 and the Office of Naval Research.







\end{document}